\documentclass[aps,pra,reprint,showpacs,superscriptaddress,groupedaddress,floatfix]{revtex4-1}
\usepackage{amsfonts,amssymb,amsmath}
\usepackage{bm}
\usepackage{graphicx}
\usepackage{hyperref} 
\usepackage{enumitem}
\setlist{nolistsep,leftmargin=*}
\providecommand{\abs}[1]{\left\lvert#1\right\rvert}
\providecommand{\norm}[1]{\left\lVert#1\right\rVert}
\providecommand{\inner}[1]{\left\langle#1\right\rangle}
\providecommand{\eps}{\varepsilon}
\providecommand{\ee}{\text{e}}
\providecommand{\oo}{\text{o}}

\providecommand{\D}{\partial}
\providecommand{\R}{\mathbb{R}}
\providecommand{\Z}{\mathbb{Z}}
\providecommand{\lamsharp}{{\lambda_{\sharp}}}
\providecommand{\thetasharp}{{\vartheta_{\sharp}}}

\newtheorem{theorem}{Theorem}
\newtheorem{corollary}{Corollary}

\begin{document}

\title[]{Photonic realization of topologically protected bound states in domain-wall waveguide arrays}

\date{\today}
\author{J.~P. Lee-Thorp$^1$,
  I.~Vuki\'{c}evi\'{c}$^1$, X.~Xu$^2$, J.~Yang$^3$, C.~L. Fefferman$^4$,  C.~W. Wong$^3$ and M.~I. Weinstein$^5$}
 \affiliation{$^1$ Department of Applied Physics and Applied Mathematics, Columbia University, New York, NY, USA;
jpl2154@columbia.edu, iv2143@columbia.edu \\
 $^2$ Department of Computer Science, Columbia University, New York, NY, USA; 
xx2153@columbia.edu\\
 $^3$ Mesoscopic Optics and Quantum Electronics Laboratory, University of California, Los Angeles, CA and
 Department of Mechanical Engineering, Columbia University, New York, NY, USA; 
yangjh@seas.ucla.edu, cheewei.wong@ucla.edu\\
$^4$ Department of Mathematics, Princeton University, Princeton, NJ, USA;
cf@math.princeton.edu\\
$^5$ Department of Applied Physics and Applied Mathematics and Department of Mathematics, Columbia University, New York, NY, USA; miw2103@columbia.edu}

\begin{abstract}
We present an analytical theory of  topologically protected photonic states for the two-dimensional Maxwell equations for a class of continuous periodic dielectric structures, modulated by a domain wall. We further numerically confirm the applicability of this theory for three-dimensional structures.
\end{abstract}

\maketitle

\section{Introduction\label{intro}}

Due to their role as vehicles for localization and transport of energy, surface modes  have long been recognized to be of central importance \cite{tamm1932uber,shockley1939surface}.
A striking  example is that of \emph{topologically protected edge states}, which were first studied in the quantum Hall effect in condensed matter physics  \cite{halperin1982quantized, wen1991gapless, TKNN:82}. Since 2005, there has been intense focus on the protected states of topological insulators \cite{Kane-Mele:05,HK:10}. More recently, there has also been growing interest in \emph{photonic} analogues of the topologically protected states observed in electronic systems \cite{raghu2008analogs, haldane-raghu-prl:08, lu2014topological}.  In particular, such states have been shown to arise  via  line-defects (edges) and time-reversal or spatial-inversion symmetry breaking in linear and nonlinear photonics, {\it e.g.} \cite{raghu2008analogs,haldane-raghu-prl:08,soljacic-etal2008,joannopoulos2011photonic,plotnik2013observation, lu2013weyl, kocaman2011zero, kocaman2009observation, jiao2006systematic}.

Theoretical arguments for the existence and robustness of topologically protected edge states have been given in terms of topological invariants associated with the band structure, {\it e.g.} Chern index or Zak phase backed by numerical simulations.
A systematic justification of the applicability of such invariants, however, appears only to have been provided  in  tight-binding models, {\it e.g.} \cite{Delplace-etal:11,Mong-Shivamoggi:11,graf2013bulk,chiu2015classification}. In the photonic setting, this corresponds to the limit of infinite medium contrast. 
We seek a general understanding of topological protection, a theory which is applicable outside of limiting regimes, such as the tight-binding or nearly-free-photon approximations.
Indeed, many physical settings of interest, including photonics in low contrast periodic structures, fall outside these regimes; see also \cite{plotnik2013observation}.

In \cite{FLW-PNAS:14,FLW-arXiv:14}, a class of one-dimensional (1D) continuous systems described  by the Schr\"odinger
equation with a domain-wall modulated (DWM) periodic potential were proved to have
 topologically protected bound states, originating in the zero-mode of an effective 1D Dirac equation, a mechanism which  
plays a role in  \cite{su1979solitons,raghu2008analogs}; see also \cite{JR:76}. In this Letter, we propose and investigate a photonic
realization of these states as topologically protected guided wave modes in a class of coupled two-dimensional (2D) waveguide arrays governed by Maxwell's equations.  No tight-binding or nearly-free-photon assumptions are made;
the results hold for bulk structures of arbitrary material contrast.

We further demonstrate,  through  \emph{ab initio} numerical simulations for the full three-dimensional (3D)  Maxwell equations, that these waveguide modes  and their properties persist for physically realistic 3D optical waveguide arrays. 
In particular, robustness against  geometric and intrinsic background thermal perturbations is demonstrated. These states have fundamentally linear dispersion and  broadband flat group velocity dispersion.

Photonic waveguide array structures are an important class of photonic crystals, providing a natural platform for basic studies of topological insulators and their photonic applications. They permit qualitative study and quantitative measurements of many phenomena \cite{bromberg2009quantum, joglekar2013optical}. 
% along the propagation length \cite{bromberg2009quantum, joglekar2013optical}. 
Recent work has explored 2D arrays in linear and nonlinear regimes, in both deterministic and random media  \cite{khanikaev2013photonic, rechtsman2013photonic,ma2015guiding,titum-disordered-fti-arXiv,zeuner-etal:14,rechtsman-etal:13,lumer-etal:13}. 

\section{Schr\"odinger setting\label{schro}}

\subsection{Guided TM Maxwell modes as eigenmodes of the Schr\"odinger equation\label{2d_tm_maxwell}}

The propagation of light in space, with coordinates $(x,y,z)$, in a dielectric medium with constant permeability and permittivity depending on only one variable, $\eps(x)$, is governed by Maxwell's equations. Time-harmonic modes, which propagate in the $z$-waveguided direction are of transverse-electric (TE) $(E_1,H,E_2)$ or transverse-magnetic (TM) $(H_1,E,H_2)$ type. We focus  on the TM case; an analogous discussion holds for TE modes.

TM modes with frequency $\omega$, which are localized in the transverse $(x)$ direction, are of the form
\begin{equation}
E(x,z,t) = e^{i\omega t} e^{\pm i \sqrt{-\mu}\ \frac{\omega}{c} z} \Psi\Big(\frac{\omega}{c}  x\Big) 
\label{guided_mode},
\end{equation} where $(\mu,\Psi(x'))$ is a solution of the Schr\"odinger eigenvalue problem (EVP) with potential $-\eps(\frac{c}{\omega}x')$, and energy $ \mu = -\frac{c^2}{\omega^2} k_z^2$: 
\begin{equation}
\left(-\frac{\D^2}{\D x'^2} - \eps\Big(\frac{c}{\omega}x'\Big) \right) \Psi(x') = \mu \Psi(x'),\ \  \Psi\in L^2(\R).
\label{schroedinger}
\end{equation} 
Here, $\eps(x')$ denotes the guided mode effective permittivity as a function of the dimensionless transverse distance $x'=\frac{\omega}{c}x$, $k_z$ is the $z$-propagation constant, and $c$ is the vacuum speed of light.  If $(\mu,\Psi)$ is an eigenpair of \eqref{schroedinger} for which $k_z^2=-(\omega/c)^2\mu>0$, then $k_z$ is real and \eqref{guided_mode} is a guided TM mode,  propagating in $z$ and localized in $x$.

\subsection{Topologically protected bound states\label{schro_summary}}

\begin{figure}
\centering
\includegraphics[width=\columnwidth]{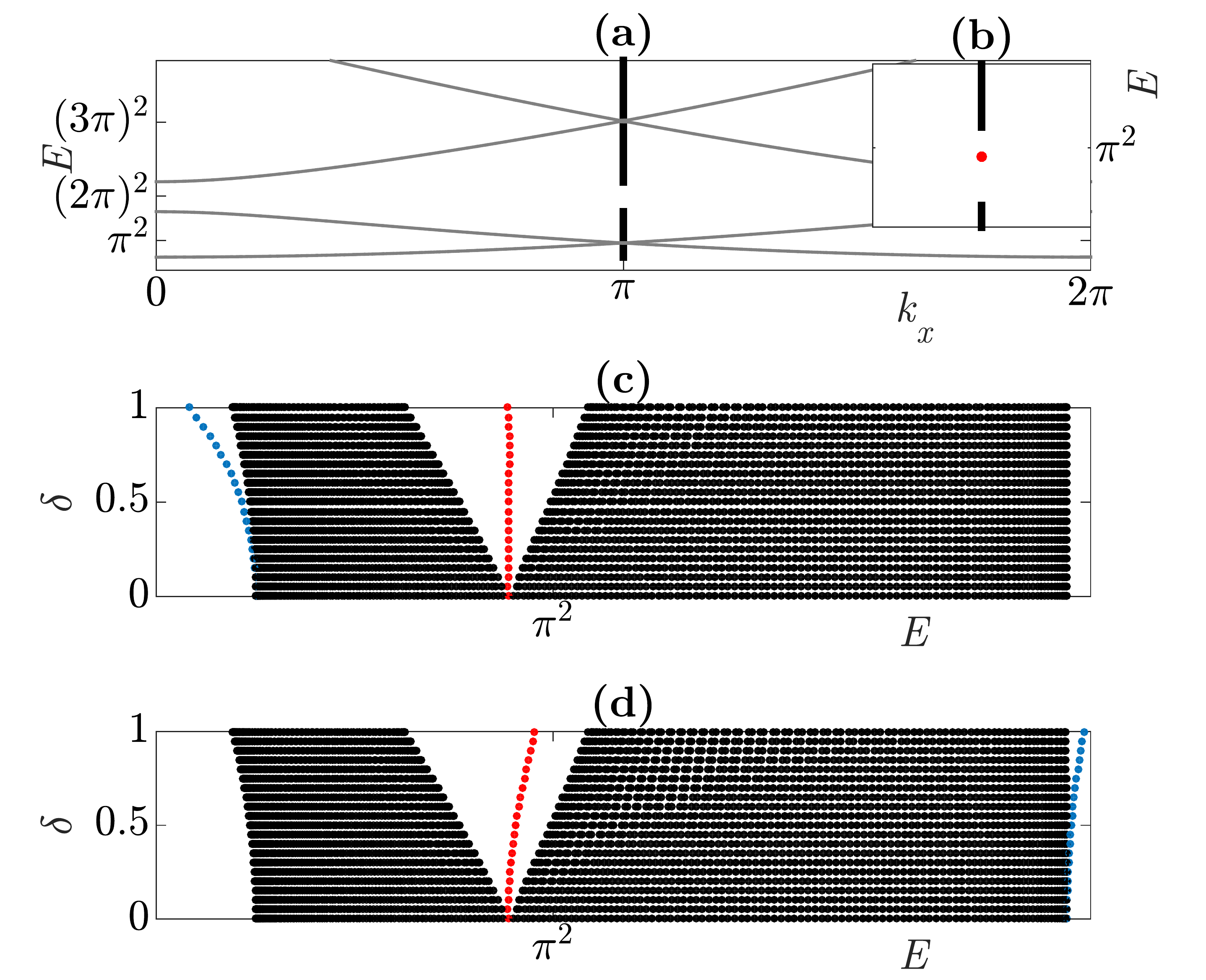}
\caption{\label{1D_Schro_theory}{
Spectra of $H^{\delta}=-\D_x^2+V_\ee(x)+\delta\kappa(\delta x)W_\oo(x)$.
{\bf (a)}: $\delta=0$. First $4$  dispersion curves $k_x\mapsto E_b(k_x)$ (gray) and continuous spectrum (black) of $H^0=-\D_x^2+V_\ee(x)$. Dirac points at $k_{x,\star}=\pi$, $E_\star\approx\pi^2, (3\pi)^2$. 
{\bf (b)}: Spectra of $H^{\delta}$ with $\kappa(X)=\tanh(X)$.
Defect mode energy, $E^\delta$  (red dot), lies in the spectral gap about  $E_\star=\pi^2$.  
{\bf (c)}: For a range of  $\delta>0$, continuous spectrum (black) of $H^\delta$ with domain wall function $\kappa(X)=5\tanh(X)+10e^{-X^2}$. Red curves are topologically protected bifurcations, seeded by effective Dirac equations. Blue curves are (nonprotected) band-edge bifurcations, seeded by effective Schr\"odinger equations.
{\bf (d)}: Spectrum of $h^\delta = H^\delta + \delta^2 G(\delta x)$, with $G(X)= 3 e^{-X^2}$ and $\kappa(X)$ as in (c).
Bifurcations from Dirac points (red) persist while those from band edges (blue) may be destroyed when subjected to localized perturbations. 
}}
\end{figure}

In \cite{FLW-PNAS:14,FLW-arXiv:14}, the authors study topologically protected bound states in 1D domain-wall modulated (DWM) Schr\"odinger Hamiltonians of the form: $H^\delta  =  -\D_x^2 + V_{\ee}(x) + \delta\kappa(\delta x) W_{\oo}(x),$ where $V_\ee$ and $W_\oo$ denote even-index and odd-index cosine series, respectively; and $\kappa(X)$ is a so-called domain wall function that satisfies $\kappa(X)\to \pm\kappa_\infty,$ $X\to\pm\infty$, with $\kappa_\infty>0$.
$H^\delta$ interpolates between \emph{dimer} structures at $x=\pm\infty$.
For $\delta=0$, $H^0$ has distinguished quasimomentum-energy pairs where its dispersion curves cross linearly; see FIG. \ref{1D_Schro_theory}(a).

\begin{theorem}[\cite{FLW-PNAS:14,FLW-arXiv:14}]\label{dirac-pt-gen}\
The Floquet-Bloch EVPs for $H^0=-\D_x^2+V_\ee(x)$: $H^0\Phi(x;k_x) = E\Phi(x;k_x)$,  $\Phi(x+1;k_x) = e^{i k_x}\Phi(x;k_x)$, parametrized by $k_x\in[0,2\pi]$, possess Dirac points. These are pairs $(k_{x,\star}=\pi,E_\star)$ with mappings $k_x\mapsto \Big(E_\pm(k_x),\Phi_\pm(x;k_x)\Big)$ such that the dispersion locus near $(k_{x,\star},E_\star)$ is given by:
$E_\pm(k)-E_\star\ =\pm\lambda_\sharp\left(k-k_{x,\star}\right)
 +\mathcal{O}\Big((k_x-k_{x,\star})^2\Big).$
Here, $\lambda_{\sharp}=2i\inner{\Phi_1,\D_x\Phi_2}_{L^2([0,1])}\neq0$ is the ``Fermi velocity''.
\end{theorem}

Associated with each Dirac point of the periodic (unmodulated) Hamiltonian, $H^0$, there is a topologically protected branch of bound states of the Schr\"odinger EVP for the DWM Hamiltonian, $H^\delta$.

\begin{theorem}[\cite{FLW-PNAS:14,FLW-arXiv:14}]
\label{bound_states}
Let $(k_{x,\star}=\pi,E_\star)$ denote a Dirac point of $H^0 = -\D_x^2+V_\ee(x)$.
Assume the (generically satisfied) non-degeneracy condition: $\lambda_\sharp\times\vartheta_\sharp\ne0$, where $\lambda_{\sharp}=2i\inner{\Phi_1,\D_x\Phi_2}_{L^2([0,1])}$ and 
$\vartheta_\sharp=\inner{\Phi_1,W_{\oo}\Phi_2}_{L^2([0,1])}$.
\begin{enumerate}
\item There exists, for small $\delta$, a family of exponentially localized eigensolutions, $\delta\mapsto(E^\delta,\Psi^\delta)$ of the EVP: 
$H^\delta\Psi^\delta=E^\delta\Psi^\delta,\ \Psi^\delta\in L^2(\R)$,
 bifurcating from energy $E_\star$ at $\delta=0$.
 \item $\Psi^\delta(x)$ is well-approximated by a slowly varying and spatially decaying modulation of the degenerate modes $\Phi_1$ and $ \Phi_2$: 
$\Psi^\delta(x) \approx\ \alpha_{\star,1}(\delta x)\Phi_1(x)+\alpha_{\star,2}(\delta x)\Phi_1(x)$.
The amplitude vector, $\alpha_\star=\left(\alpha_{\star,1},\alpha_{\star,2}\right)$, 
 is a zero mode of a 1D Dirac operator: 
$\mathcal{D}\equiv i\lamsharp\sigma_3\D_{X}+\thetasharp\kappa(X)\sigma_1$, where $\sigma_1$ and $\sigma_3$ are the standard Pauli matrices.
  \item This zero-energy state of  $\mathcal{D}$ persists for arbitrary spatially localized perturbations of the domain wall, $\kappa(X)$, and hence the bifurcation is topologically protected.
\end{enumerate}
\end{theorem}

FIG. \ref{1D_Schro_theory}(b)-(c) illustrate aspects of Theorem \ref{bound_states}  for a  range of $\delta$-values. In particular, the first branch of mid-gap eigenmodes (red) bifurcates into $\mathcal{I}_\delta\equiv(E_\star-\delta|\kappa_\infty\vartheta_\sharp|,E_\star+\delta|\kappa_\infty\vartheta_\sharp|)$.

\subsection{Protected vs non-protected modes\label{non_protected}}

To contrast the protected modes of DWM structures with conventional non-protected modes of periodic structures with defects which arise via bifurcations from a band-edge energy, $E_b(k_{\rm edge})$, consider the Hamiltonian 
$h^\delta =-\D_x^2+V_\ee(x)+\delta\kappa(\delta x)W_\oo(x) + \delta^2 G(\delta x)$, 
which incorporates both a domain-wall induced phase defect ($\delta\kappa(\delta x)W_\oo(x)$) and a localized defect ($\delta^2 G(\delta x)$); scalings are chosen so that both effects are of comparable size.
Band-edge bifurcations are seeded by bound states of the effective Schr\"odinger operator:
$h_{\rm b, eff} =-(2{\rm m}_{b,{\rm eff}})^{-1}\D_X^2+Q_{b,{\rm eff}}(X)$, 
with effective mass \cite{kittel:96}  ${\rm m}_{b,{\rm eff}}^{-1}=E_b''(k_{\rm edge})$  and effective potential $Q_{b,{\rm eff}}(X)= c_b \ (\kappa_\infty^2-\kappa^2(X)) + G(X)$, $c_b>0$.
While the zero-energy eigenstate of $\mathcal{D}$, which induces the bifurcation of bound states from a Dirac point \cite{shockley1939surface},
 persists under arbitrary (even large) localized perturbations of the domain wall, $\kappa(X)$, the localized states of $h_{\rm b, eff}$, and therefore their induced branch of bound states \cite{tamm1932uber}, are not stable to arbitrary localized perturbations, $G(X)$ \cite{hoefer-weinstein:11,dvw-cms:15}.
  
In FIG. \ref{1D_Schro_theory}(c), where $G\equiv0$, the domain wall induces bifurcations from Dirac points as well as from band edges.
In FIG. \ref{1D_Schro_theory}(d), a non-zero localized perturbation, $G(X)$, has destroyed the lower band edge bifurcation; the edge bifurcation can be removed by smooth deformation of $h^\delta$ while the Dirac point bifurcation is topologically protected. 

\begin{figure}
\centering
\includegraphics[width=\columnwidth]{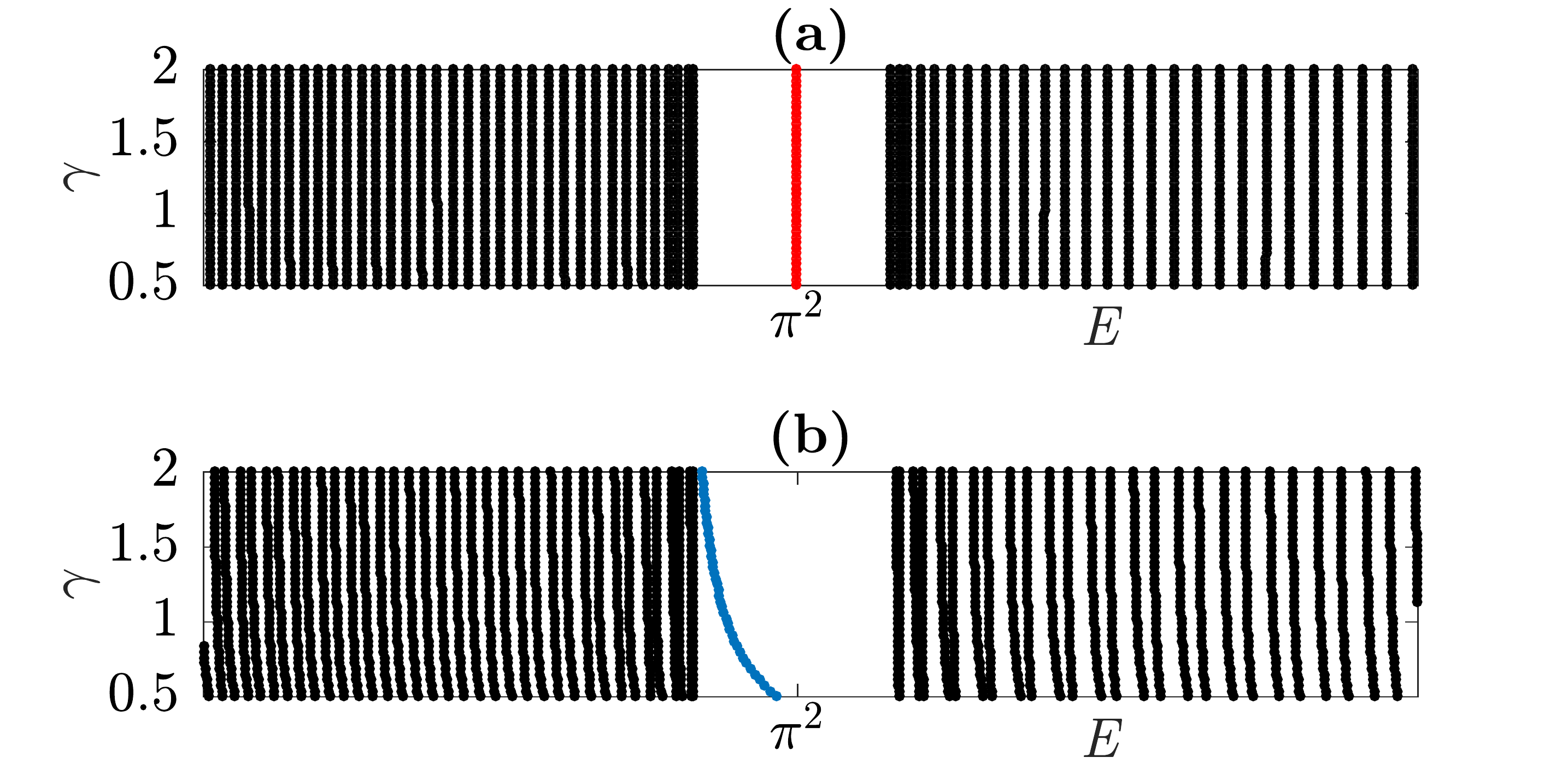}
\caption{\label{1D_Schro_robustness}{
Greater robustness of operating frequency of protected modes when compared with conventional non-protected modes, illustrated by the spectra of $h^{\delta}_\gamma =-\D_x^2+V_\ee(x)+\delta\kappa(\gamma \delta x)W_\oo(x) + \delta^2 G(\gamma \delta x)$ near first spectral gap.
% Spectra of $h_\gamma =-\D_x^2+V_\ee(x)+\kappa(\gamma x)W_\oo(x) + G(\gamma x)$ near first spectral gap.
{\bf (a)}: Protected mode; $G\equiv0$ and $\kappa(X)=\tanh(X)$. 
{\bf (b)}: Non-protected mode; $\kappa\equiv1$ and $G(X)=3e^{-X^2}$.
}}
\end{figure}

The protected states of DWM structures also exhibit a remarkable robustness of their eigenvalues (operating or working frequencies) when compared with those of conventional structures with localized defects. To illustrate this, we  
extend the previous model to a Hamiltonian $h^{\delta}_\gamma =-\D_x^2+V_\ee(x)+\delta\kappa(\gamma \delta x)W_\oo(x) + \delta^2 G(\gamma \delta x)$, with defect mode eigenvalue, $E_\delta^\gamma$,  where changing $\gamma$ varies the phase defect's transition region width (where $\kappa$ transitions between positive to negative values) and the area of the localized defect, $G$. The potential in $h_\gamma^\delta$ is assumed to be smooth. For $\gamma\approx1$ and $0<\delta\ll1$, $E_\delta^\gamma- E_\delta^{\gamma=1}$ is of the order 
$(\Delta E)_\kappa=(\gamma-1) \delta\times \int  (\delta x) \kappa'(\delta x) W_\oo(x) |\Psi_\kappa^{\delta}|^2 dx$
for the protected mode (DWM $\kappa$, $G\equiv0$), and of the order
$(\Delta E)_G=(\gamma-1)\delta^2\times \int (\delta x) G'(\delta x) |\Psi_G^{\delta}|^2 dx$
for the non-protected mode ($\kappa\equiv1$, localized $G\neq0$).

For $\delta\ll1$, we have  $\Psi_{\kappa}^{\delta}, \Psi_{G}^{\delta} \approx \delta^{1\over2}\ e^{-c|\delta x|},\ c>0$; see Theorem \ref{bound_states} and \cite{dvw-cms:15,FLW-PNAS:14}. Therefore,  using the rapid oscillations of $W_\oo(y/\delta)$, we have 
$(\Delta E)_\kappa \approx (\gamma-1)\int F(y)W_\oo(y/\delta) dy = \mathcal{O}\left((\gamma-1)\delta^{M}\right)$, 
for all $M\in\Z^+$ where  $F(y) =  y\kappa'(y)e^{-2|y|}$. On the other hand, a direct computation gives $(\Delta E)_G = \mathcal{O}\left((\gamma-1)\delta\right)$. Hence, for small $\delta$ and $\gamma\approx1$, we have $|(\Delta E)_\kappa|\ll |(\Delta E)_G|$.
Numerical simulations for larger $\delta$ and $|\gamma-1|$ further corroborate the strong robustness of the operating frequency of topologically protected modes; see FIG. \ref{1D_Schro_robustness}.

\section{Maxwell setting\label{maxwell}}

\subsection{2D topologically protected, guided TM modes\label{2d_maxwell_protected}}

Returning to the Maxwell setting, Theorem \ref{bound_states} and \eqref{guided_mode} imply:

\begin{corollary} \label{TM}
Maxwell's equations exhibit topologically protected, transversely localized, guided TM modes. 
\end{corollary}

\begin{figure}
\centering
\includegraphics[width=\columnwidth]{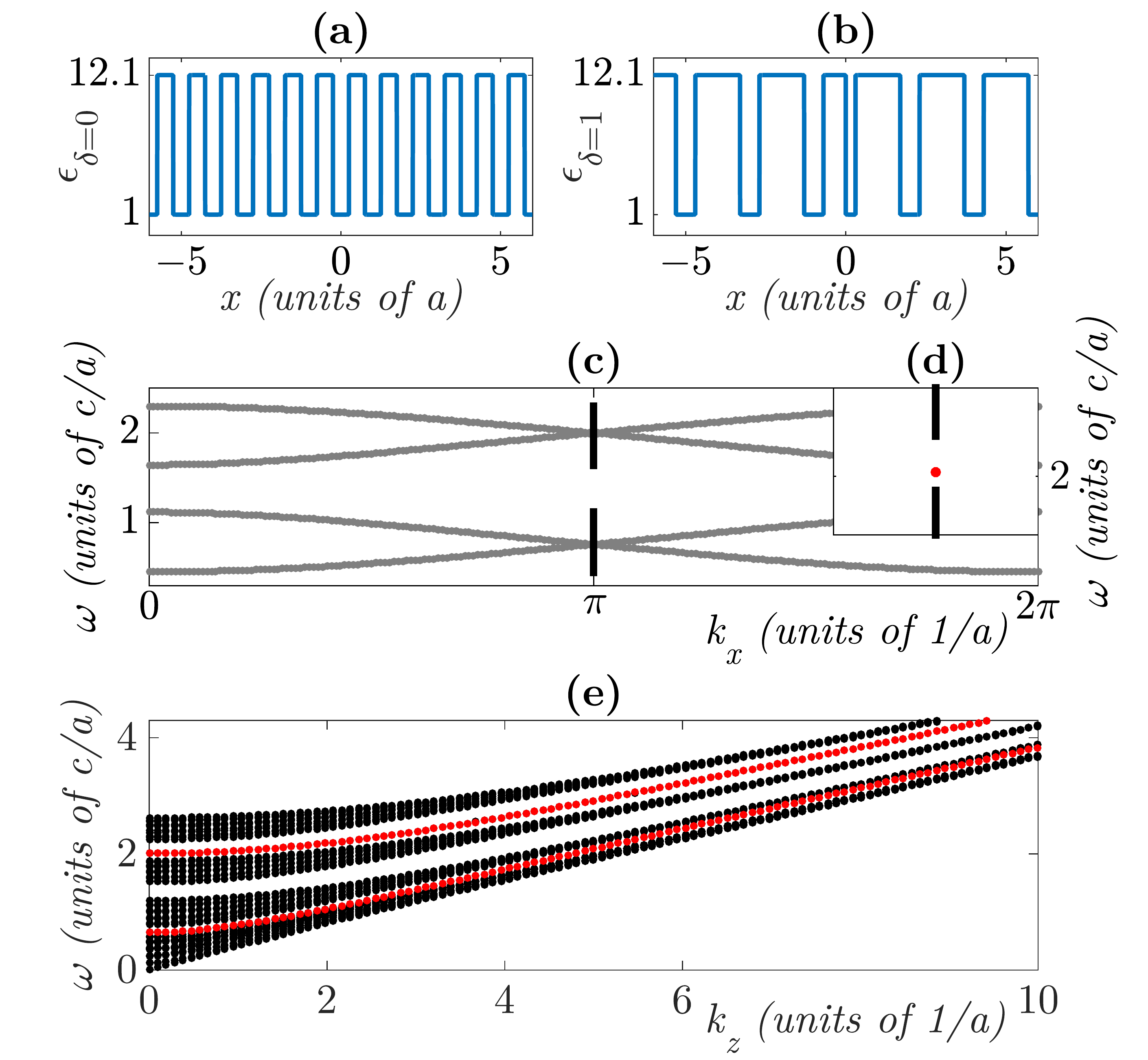}
\caption{\label{2D_Maxwell}{
2D Maxwell problem.
{\bf (a)}: Periodic ($\delta=0$) 2D silicon-air waveguide array with lattice constant (period) $a$.
{\bf (b)}: DWM ($\delta=1$) 2D silicon-air  array. 
Left and right asymptotic $2a$-periodic structures are phase-shifted relative to one another.
Modulated structures consisting of $5$ waveguide channels ($3$  shown here) on either side of the defect region were used for numerical computations.
{\bf (c)}: TM dispersion curves, $\omega(k_x)$ (gray), and frequency spectra (black) of the first few TM bands of the periodic 2D waveguide array ($\delta=0$), with $k_z=2\pi/10$ fixed. Linear band crossings (Dirac points) occur at $k_x=\pi/a$.
{\bf (d)}: For the 2D DWM waveguide array ($\delta=1$), bound state energies (red dot) bifurcate into spectral gaps about Dirac points of the periodic (unmodulated) structure.
{\bf (e)}: Dispersion curves, $\omega(k_z)$, of the first few TM bands for the 2D DWM waveguide.
Eigenvalue curves (red), corresponding to guided modes, lie in between bands of continuous spectrum (black).}}
\end{figure}

We realize these topologically robust TM modes in a photonic waveguide array by constructing a coupled 2D silicon-air waveguide array profile, $\eps_\delta(x)$, corresponding to a potential, $\mathcal{U}_\delta(x)$. 
 We construct  $\mathcal{U}_\delta(x)$ to be the two-valued approximation of a choice of
  $V_\ee(x)+\delta\kappa(\delta x)W_\oo(x)$, scaled 
  to match the effective permittivities of silicon ($12.1$) and air ($1$).
$\mathcal{U}_{\delta=1}(x)$ is the DWM  structure and $\mathcal{U}_{\delta=0}(x)$ is the periodic potential having Dirac points. 
Finally, we set $\eps_\delta(\frac{c}{\omega}x)=-\mathcal{U}_\delta(x)$; see \eqref{schroedinger}.
The 2D DWM waveguide array is shown in FIG. \ref{2D_Maxwell}(b). Although the relative length scale between $\mathcal{U}_\delta(x)$ and $\eps_\delta(x)$ is important, the computations may be carried out at arbitrary length scales due to the scale invariance of Maxwell's equations \cite{joannopoulos2011photonic}.

Since the numerically computed lowest energy gap mode of the Schr\"odinger equation with the DWM potential, $\mathcal{U}_{\delta=1}(x)$, satisfies $\mu<0$, $k_z=\pm(\omega/c)\sqrt{-\mu}$ is real and the corresponding TM mode in \eqref{guided_mode} is guided.

The dispersion curves, 
 $k_x\mapsto\omega(k_x)$, for TM modes of the 2D  periodic (unmodulated) structure, $\eps_{\delta=0}(x)$, are displayed in FIG. \ref{2D_Maxwell}(c). 
These curves are computed by solving the 2D Maxwell equations for $\omega$ at discrete $k_x$ values, using MPB \cite{Johnson2001:mpb}.
The dispersion curves, 
 $k_z\mapsto\omega(k_z)$, for the DWM waveguide, $\eps_{\delta=1}(x)$, are displayed in FIG. \ref{2D_Maxwell}(e). In analogy with the Schr\"odinger setting, robust guided-modes
are observed in gaps about Dirac points at quasi-momentum $k_x=\pi/a$, where $a$ is the lattice constant; see FIG. \ref{1D_Schro_theory}(a)-(b).
Qualitatively similar dispersion curves are observed for TE modes.

\begin{figure}
\centering
\includegraphics[width=\columnwidth]{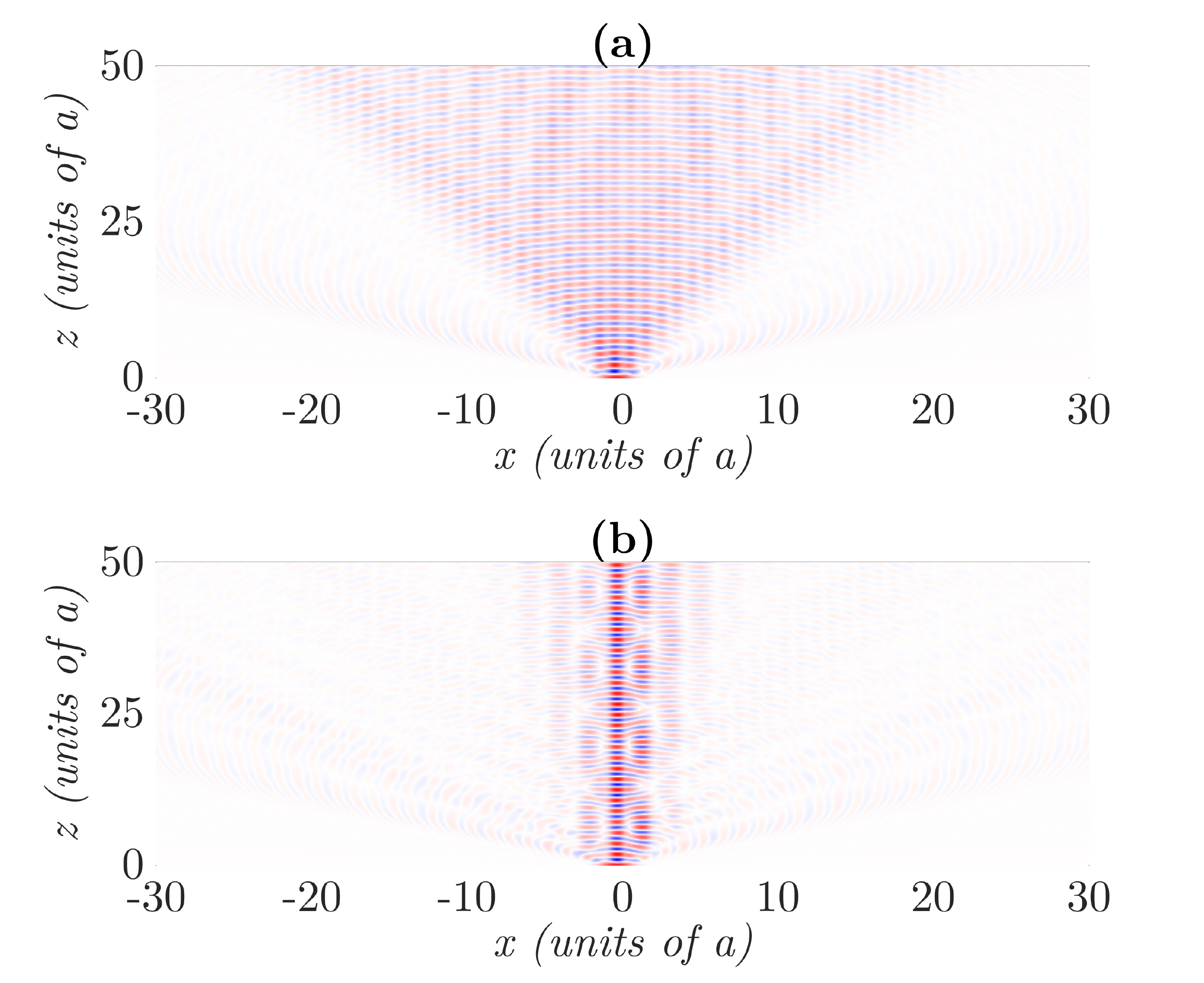}
\caption{\label{2D_propagating}{
On-axis ($z$) propagation of electric field, $E(x,z)$, through {\bf (a)} periodic and {\bf (b)} DWM 2D waveguide arrays.
}}
\end{figure}

We numerically  study on-axis $z$-propagation in the DWM 2D waveguide of a Gaussian wave-packet, using the approximate paraxial Schr\"odinger equation \cite{lifante2003front,kawano2004introduction}.
For the periodic array, the energy of the launched packet quickly delocalizes (FIG. \ref{2D_propagating}(a)). In contrast,  in the DWM structure, a topologically protected mode is excited and confines most of the energy on-axis (FIG. \ref{2D_propagating}(b)). 
A quantitative comparison of localization properties of the DWM and periodic structures is obtained by measuring, $\xi$, the fraction of power  remaining within (roughly) the width of the profile of the stationary localized mode of the DWM structure at $z=50a$: $\xi= \norm{E(x,z=50a)}_{L^2([-5a,5a])}^2/\norm{E(x,z=0)}_{L^2([-30a,30a])}^2$. 
The DWM structure achieves significantly more localization:
$ \xi_{\rm DWM} = 0.92 \ \text{and} \ \xi_{\rm periodic} = 0.35$. 

\subsection{3D topologically protected, guided TM modes\label{3d_maxwell_protected}}

\begin{figure}
\centering
\includegraphics[width=\columnwidth]{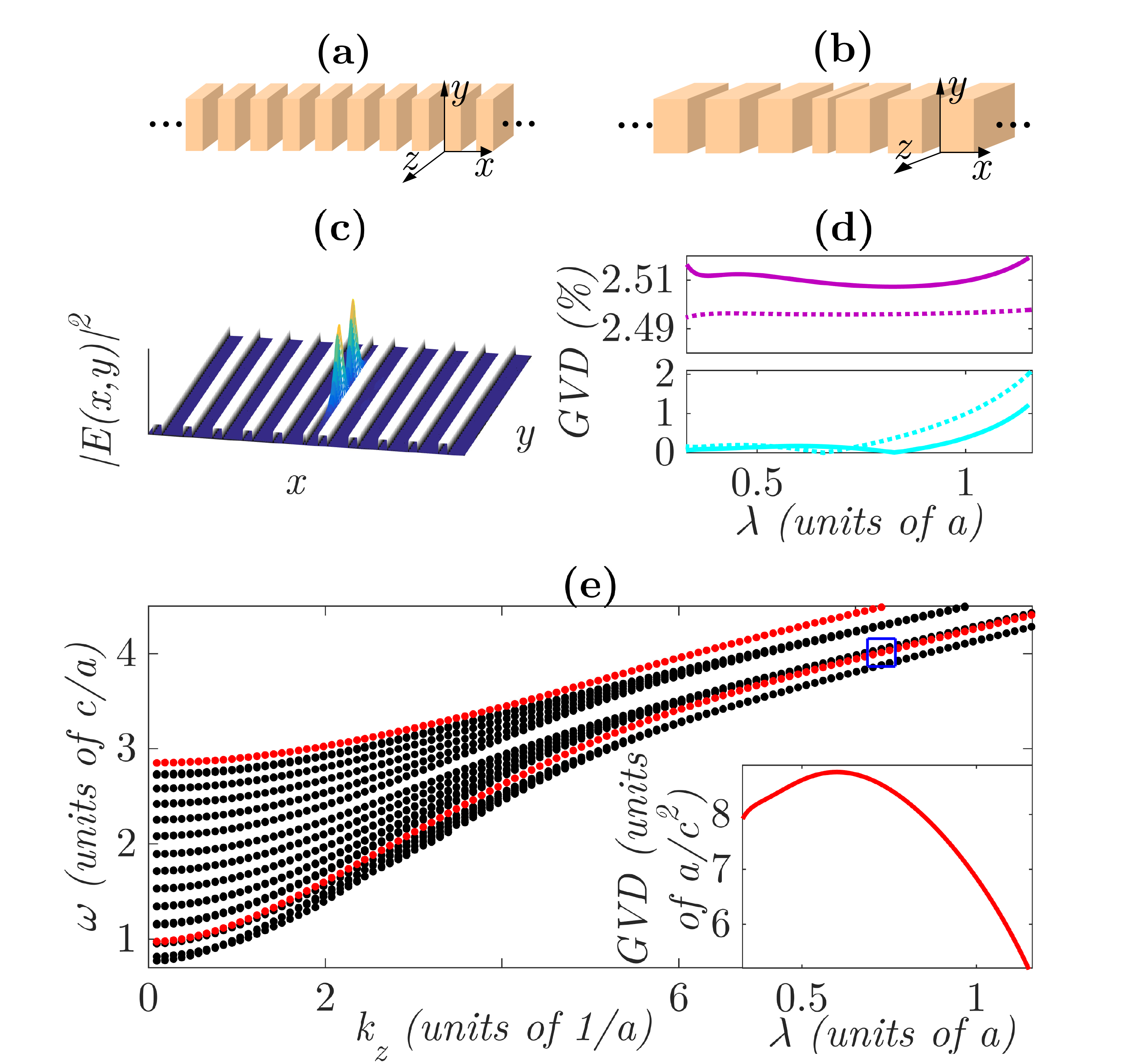}
\caption{\label{3D_Maxwell}{
3D Maxwell problem.
{\bf (a)}: Periodic ($\delta=0$) 3D silicon waveguide with lattice constant ($x$-direction period) $a$ and height ($y$-direction) $0.25 a$, and silicon dioxide cladding (not shown). Height exaggerated for clarity.
{\bf (b)}: DWM ($\delta=1$) 3D silicon waveguide with height $0.25 a$, and silicon dioxide cladding. 
Computations were carried out on DWM structures with $5$ waveguide channels on either side of the defect region.
{\bf (c)}: Electric field $\abs{E}^2$ of the localized mode with $\omega=4.0c/a^{}$ and $k_z=8.3/a^{}$ (blue square in (e)), over a fixed $z$-slice of the DWM waveguide.
{\bf (d)}: Mean relative deviation of GVD vs. wavelength $\lambda$, of the DWM (solid curves) and regular channel (dotted curves) waveguides when subjected to thermal (purple) and geometric (cyan) perturbations.
{\bf (e)}: Dispersion curves, $\omega(k_z)$, of the first few TM bands of the (unperturbed) 3D DWM waveguide. Inset shows the associated GVD of lower frequency protected guided modes.
}}
\end{figure}

The theory summarized in Theorem \ref{bound_states} and Corollary \ref{TM} is exact and rigorous for a 2D waveguide with effective permittivity satisfying $\eps=\eps(x)$. The physical 3D DWM waveguide is a finite height ($y$) truncation of the 2D structure. 
Numerical simulations (below) show a long lived 3D state, localized in the $x$-direction even for small height truncations.  A schematic of a 3D silicon waveguide (effective permittivity $12.1$) with height $0.25 a$ and silicon-dioxide cladding (effective permittivity $2.085$) is shown in FIG. \ref{3D_Maxwell}(b).

Numerically computed dispersion curves:  $k_z\mapsto\omega(k_z)$, for TM modes of the 3D DWM waveguide are displayed in FIG. \ref{3D_Maxwell}(e). From $\omega(k_z)$, we compute the group velocity dispersion (GVD), $\partial^2k_z/\partial\omega^2$; see figure inset.

Corollary \ref{TM} ensures that the 2D TM-modes, which bifurcate from Dirac points, are topologically stable. 
We confirm the existence and persistence of their 3D counterparts through a computational study of two classes of physical perturbations:
(i) changes in the effective permittivity due to thermal fluctuations; and (ii) geometric perturbations in the lattice constants (waveguide channels) of the array due to fabrication imperfections, both  $10\%$ perturbations, randomly sampled from a uniform distribution.
FIG. \ref{3D_Maxwell}(d)  (solid) displays the mean deviation of the GVD curves of perturbed waveguides relative to the unperturbed (DWM) waveguide, calculated from one hundred independent simulations. Also plotted (dotted), for reference, are the corresponding mean relative deviation of the GVD curves for a regular channel silicon-silicon dioxide waveguide (of height $0.25 a$ and channel width $0.45a$).
GVD robustness of topologically protected DWM guided modes, with respect to thermal and geometric perturbations, is comparable to that for regular channel modes.

Fixing $\omega=4.0c/a^{}$, we also observe that the mode itself remains localized against the perturbations, as measured by the perturbed effective mode area statistics \cite{agrawal2007nonlinear}:
$A_{\rm thermal}=0.92\pm0.09 \text{ and } A_{\rm geometric}=0.9\pm0.4.$
Here the perturbed mode areas are normalized against the unperturbed mode area, and reported as means with corresponding standard deviations, calculated from fifty independent simulations.

\section{Conclusion\label{conclusion}}

Summarizing, building upon the theory of  \cite{FLW-PNAS:14,FLW-arXiv:14}, we have demonstrated the bifurcation of highly robust guided TM modes for Maxwell's equations governing a class of DWM photonic waveguide arrays in 2D.  
Our findings are corroborated by full Maxwell simulations of physically realistic 3D structures, derived from our 2D model.
In contrast to the guided wave modes of conventional waveguides, DWM modes are robust to large localized perturbations while having comparable GVD robustness characteristics. 
These topologically protected states may be well-suited for chip-scale nanofabrication of semiconductor waveguides for communications and frequency source generation.

\begin{acknowledgments}
The authors thank C. Wilson and M. Spiegelman at Columbia LDEO for providing valuable computing resources,  E. D. Kinigstein and J. F. McMillan for helpful discussions, and the referees for many stimulating insightful comments.
This work was supported, in part, by NSF Grants DMS-1265524 (C.L.F.), DMS-1008855 and DMS-1412560 (J.P.L.-T., M.I.W.), CMMI-1520952 (C.W.W., J.Y.), (IGERT) DGE-1069420 (J.P.L.-T., M.I.W., C.W.W., J.Y.); ONR Grant N00014-14-1-0041 (C.W.W., J.Y.); and the Simons Foundation Grant 376319 (M.I.W.).
\end{acknowledgments}

\bibliography{experiment-notes}

\end{document}